\documentclass[aps,prl,a4paper,preprintnumbers,twocolumn,floatfix,showpacs]{revtex4}
\usepackage{amsmath,amsfonts,amssymb,mathrsfs,color}
\usepackage{graphicx}
\usepackage{subfigure}
\usepackage{enumerate}
\def\be{\begin{equation}}
\def\ee{\end{equation}}
\def\bea{\begin{eqnarray}}
\def\eea{\end{eqnarray}}
\def\nn{\nonumber}

\begin{document}
\title{Black hole hair in generalized scalar-tensor gravity}
\author{Thomas P. Sotiriou$^{1}$  and Shuang-Yong Zhou$^{2}$}
\affiliation{$^{1}$School of Mathematical Sciences \& School of Physics and Astronomy, University of Nottingham, University Park, Nottingham, NG7 2RD, UK
\\$^2$SISSA, Via Bonomea 265, 34136, Trieste, Italy {\rm and} INFN, Sezione di Trieste, Italy}
\begin{abstract}
The most general action for a scalar field coupled to gravity that leads to second order field equations for both the metric and the scalar --- Horndeski's theory --- is considered, with the extra assumption that the scalar satisfies shift symmetry. We show that in such theories the scalar field is forced to have a nontrivial configuration in black hole spacetimes, unless one carefully tunes away a linear coupling with the Gauss--Bonnet invariant. Hence, black holes for generic theories in this class will have hair. This contradicts a recent no-hair theorem, which seems to have overlooked the presence of this coupling.
\end{abstract}
\maketitle

In general relativity, black hole spacetimes are described by the Kerr metric, so long as they are stationary, asymptotically flat, and devoid of any matter in their surroundings \cite{Hawking:1971vc}. Stationarity is a reasonable assumption for black holes that are thought to be quiescent as endpoints of gravitational collapse. Astrophysical black holes are certainly not asymptotically flat, but one can invoke separation of scales in order to argue that the cosmological background should not seriously affect local physics and hence the structure of black holes.  Finally, black holes can also carry an electromagnetic charge in the presence of an electromagnetic field. It has been conjectured that  they cannot carry any other charges, which are colloquially referred to as hair \cite{wheeler}\footnote{Hairy black hole solutions do exist in Einstein--Yang--Mills theory, see Ref.~\cite{Volkov:1998cc} for a review and references therein.}. The no-hair conjecture was inspired by the uniqueness theorems for black hole solutions in general relativity \cite{israel1, israel2, Carter:1971zc, Wald:1971iw}.

Hawking has proven that black holes cannot carry scalar charge, provided that the scalar couples to the metric minimally or as described by Brans--Dicke theory \cite{hawking2}. This result has been generalised to standard scalar-tensor theories \cite{Sotiriou:2011dz} (see also earlier work by Bekenstein with the extra assumption of spherical symmetry \cite{Bekenstein:1995un,beken}).

All of these proofs actually demonstrate that the scalar has to be constant in a black hole spacetime, which is a stronger statement. Indeed, in principle, the scalar could have a nontrivial configuration without the black hole carrying an extra (independent) charge. This is sometime referred to as ``hair of the second kind''. The distinction is important if one is interested in the number of parameters that fully characterise the spacetime. But, a nontrivial configuration of the scalar is usually enough to imply that the black hole spacetime will not be a solution to Einstein's equations in vacuum, and hence it differs from the black holes of general relativity.

The known proofs do not apply to theories with more general coupling between the metric and the scalar, or derivative self-interactions of the scalar. Hence, they do not cover the most general scalar-tensor theory that leads to second-order field equations, known as Horndeski theory \cite{Horndeski:1974wa}. Restricting attention to theories  with second order field equations is justified, as higher order derivative models are generically plagued by the Ostrogradski instability \cite{ostro}. Models that belong to this class have lately received a lot of attention in cosmology, under the name {\em generalised Galileons} \cite{Deffayet:2009mn} (see also Ref.~\cite{Deffayet:2013lga} for a recent review).

If black holes have hair in these theories,  they could perhaps be used to indirectly detect the presence of a scalar field. The equivalence principle dictates that the matter should couple minimally to the metric and that it should not couple to the scalar field. This implies that direct detection in matter experiments is not promising. However, a non-trivial configuration for the scalar field would lead to a black hole solution that deviates from that of general relativity. The deviation could agree with the prediction of a certain model, and, in principle, accurate modelling of the spacetime could act a probe of the coupling of the scalar field to gravity and itself. The presence of scalar hair can also have bearing on the thermodynamical aspects of black holes in scalar-tensor theories.
It is, hence, quite important to understand whether black holes can have nontrivial scalar configurations in the most general scalar-tensor theory.

Progress in this direction was recently made in Ref.~\cite{Hui:2012qt}. It was argued there that vacuum, static, spherically symmetric, asymptotically flat black holes have no hair in the most general scalar-tensor theory that leads to second order field equations, provided that the scalar exhibits shift symmetry, {\em i.e.},~symmetry under $\phi\to \phi+{\rm constant}$.
The most general Lagrangian with these properties is the following \cite{us}
\begin{eqnarray}
\label{lagr}
 {\cal L}&=&K(X) -G_3(X)\Box\phi + G_{4}(X)R\nonumber\\&&
+G_{4X}\left[
\left(\Box\phi\right)^2-\left(\nabla_\mu\nabla_\nu\phi\right)^2
\right]\nonumber
\\
&&+G_5(X) G_{\mu\nu}\nabla^\mu\nabla^\nu\phi
-\frac{G_{5X}}{6}\Bigl[
\left(\Box\phi\right)^3
\nonumber\\&&
-3\left(\Box\phi\right)\left(\nabla_\mu\nabla_\nu\phi\right)^2
+2\left(\nabla_\mu\nabla_\nu\phi\right)^3
\Bigr],\,\,\,\,\,\,
\end{eqnarray}
where $K$, $G_3$, $G_4$, and $G_5$ are generic functions of $X:=-\partial_\mu\phi \partial^\mu\phi/2$,  $G_{iX}\equiv \partial G_i/\partial X$, $\nabla_\mu$ is the covariant derivative associated with the metric $g_{\mu\nu}$,  $\Box \equiv \nabla^\mu\nabla_\mu$, $\left(\nabla_\mu\nabla_\nu\phi\right)^2  \equiv \nabla_\mu\nabla^\nu \phi\nabla_\nu\nabla^\mu\phi$, $\left(\nabla_\mu\nabla_\nu\phi\right)^3 \equiv \nabla_\mu\nabla^\rho \phi\nabla_\rho\nabla^\nu\phi\nabla_\nu\nabla^\mu\phi$, and $R$ and $G_{\mu\nu}$ are the corresponding Ricci scalar and Einstein tensor respectively.  The class of scalar-tensor theories in which the scalar enjoys shift symmetry is an interesting one, as the scalar is protected from acquiring a mass by radiative corrections.

In what follows we will briefly review the no-hair proof of Ref.~\cite{Hui:2012qt} and we will show that it can be straightforwardly extended to slowly rotating black holes. However, we will also scrutinise its assumptions and we will uncover a hidden assumption that is not generically satisfied by the Lagrangian in Eq.~(\ref{lagr}), unless one fine-tunes away a certain combination of terms. In fact, generic theories of this type have hairy black hole solutions.

The equation of motion of the scalar in a theory described by Eq.~(\ref{lagr}) can be written as a  conservation equation for the Noether current $J^\mu$ associated with the shift symmetry $\phi \to \phi +$constant
\be
\label{eqphi}
\nabla_\mu J^\mu=0\,.
\ee
 Assuming that the metric is static and spherically symmetric, one can make the ansatz 
\be
ds^2=-f(\rho)dt^2+f(\rho)^{-1}d\rho^2+r^2(\rho)d\Omega^2
\ee
without loss of generality. The exact form of $J^\mu$ will be discussed shortly.
 The proof laid out in Ref.~\cite{Hui:2012qt} can be split into four steps. In the first step one argues that, if the scalar respects the symmetries of the metric,  so that $\phi=\phi(\rho)$, then the only non-vanishing component of $J^\mu$ in this coordinate system should be $J^\rho$. The angular components have to vanish because of spherical symmetry, and the $J^t$ component has to vanish because otherwise it would select a preferred time direction. The second step of the proof is to show that $J^\rho$ has to vanish on the horizon of a black hole. The Killing vector associated with time translations should become null at the horizon and in this coordinate system its norm is equal to $f$. So, $f$ should vanish at the horizon. If $J^\mu J_\mu= (J^\rho)^2/f$ is to remain finite, then $J^\rho$ must be zero at the horizon. The third step involves Eq.~(\ref{eqphi}), which can now be trivially integrated to give $r^2(\rho) J^\rho=\;$constant. But, $r^2$ remains finite at the horizon as a measure of the area of constant-$\rho$ spheres. This implies that $J^\rho$ has to vanish everywhere. The fourth and final step is to argue that $J^\rho=0$ implies $\phi=$constant,  and, therefore, the metric will have to satisfy Einstein's equations in vacuum (assuming that $G_4(0)=1$). 
  
 This last step is the trickiest one, as it relies on the actual dependence of the current on $\phi$ and its derivatives. It is argued in Ref.~\cite{Hui:2012qt} that $J^\rho$ should be of the form
 \be
 \label{Jrho}
J^\rho=f\,\partial_\rho\phi F(\partial_\rho\phi\, ; g, \partial_\rho g, \partial_\rho\partial_\rho g)\,,
\ee
where $F$ is some unspecified function. It is then claimed that $F$ will asymptote to a non-zero constant at spatial infinity if one imposes the minimal requirement that the theory will have a standard canonical kinetic term in the weak field regime. Asymptotic flatness requires $f\to1$ and $\phi'\to0$ at infinity. But if one then tries to go  to some smaller radius continuously,  $F$ and $f$ should remain non-zero, which implies that $\phi'$ has to vanish everywhere.

It is this last step of the proof  that we will contest 
 and, in particular, the functional dependence of $J^\mu$ on $\phi$ and its derivatives.  If the scalar respects the symmetries of the metric, then $\phi=\phi(r)$. Adopting a more conventional coordinate system with $r$ as the areal radius, the metric can take the form
\begin{eqnarray}
\label{metric}
ds^2&=&-R(r)dt^2+S(r)d\rho^2+r^2(d\theta^2+\sin^2\theta d\varphi^2)\,.
\end{eqnarray}
Using this ansatz  one can get the explicit form of the Noether current associated with shift symmetry:
\bea
\label{sphJ}
J^r&=&
-\frac{\phi'}{S}K_X + \frac{r\phi'^2 R' + 4R\phi'^2}{2rR S^2}G_{3X} \nn\\&&
+\frac{2R\phi' -2RS\phi' + 2r\phi' R'}{r^2R S^2}G_{4X} 
\nn\\
&&
- \frac{2R\phi'^3 +2r\,\phi'^3R' }{r^2R S^3}G_{4XX} 
\nn\\&&+\frac{S\,\phi'^2 R' - 3\phi'^2 R'}{2r^2R S^3}G_{5X} 
+\frac{\phi'^4 R' }{2r^2R S^4}G_{5XX}\,,
\eea
where a prime denotes differentiation with respect to $r$.
Every term does appear to depend at least linearly on $\phi'$, as required in Ref.~\cite{Hui:2012qt}. Additionally,  assuming that $K$ has a piece linear in $X$ so that in the weak field limit the standard canonical kinetic term is present in the action, the current does seem to be asymptotically proportional to a constant times $\phi'$: As $r\to \infty$, asymptotic flatness requires that $R,S\to 1$ and $R',S'\to 0$ and the terms that contain $G_i$ appear to vanish.
So, all of the requirements on Ref.~\cite{Hui:2012qt} seem to be justified.

One potential loophole could be to consider theories where $G_{i}$ or their derivatives with respect to $X$ have poles at $X\to 0$, as $X=-\phi'^2/2$. However, such theories will not, in general, admit solutions in which $\phi=\;$constant everywhere, as this would make the current diverge. As such, they do not fall under the purview of the proof in the first place. Moreover, in general, such theories would be unlikely to admit Lorentz-symmetric vacua, as the scalar would always be forced to be in a nontrivial configuration. There is an exception, though: suppose that the $G_i$ and their derivatives are such, so that they contain exactly the right negative powers of $X$ in order for $J^r$ to not have a pole at $X=0$ but instead have a piece that is $\phi$-independent. 

In order to show that this is possible, it is actually easier to go back to the action. What we are requesting is that the field equation of the scalar contains a term that does not depend on the scalar itself. The corresponding term in the Lagrangian should then be linear in the scalar, {\em i.e.}~of the form $\phi A[g]$ up to a total divergence, where $A[g]$ is a generally covariant scalar constructed from the metric and its derivatives. On the other hand, shift symmetry implies that $A$ itself should be a total divergence. We also want the term $\phi A$ in the Lagrangian to lead to a contribution to the field equations with no more than second order derivatives when varied with respect to both the scalar and the metric. There is only one choice that actually satisfies all requirements: $A={\cal G}\equiv R^{\mu\nu\lambda\kappa}R_{\mu\nu\lambda\kappa}-4 R^{\mu\nu}R_{\mu\nu}+R^2$, {\em i.e.}~$\phi$ has to have a linear coupling with the Gauss--Bonnet invariant.

Indeed, consider the theory
\be
\label{gbaction}
S=\frac{M_p^2}{2}\int d^4x \sqrt{-g} \left( R - \frac{1}{2}\partial_\mu \phi \partial^\mu \phi +\alpha \phi \mathcal{G} \right)\,,
\ee
where $\alpha$ is a coupling constant and $M_p$ is the reduced Planck mass. Variation with respect to $\phi$ yields
\be
\label{eqphi2}
\Box \phi+\alpha{\cal G}= \nabla_\mu (\nabla^\mu\phi +\alpha \bar{G}^\mu)=0\,
\ee
where $\bar{G}^\mu$ is implicitly defined by ${\cal G}=\nabla_\mu \bar{G}^\mu$. ${\cal G}$ vanishes {\em only} in flat space, which implies that $\phi$ cannot be constant everywhere for any other spacetime, including black holes. Although unlikely, it is not {\em a priori} inconceivable that  black hole solutions do not exist at all in this model. This is not the case and we will provide explicit
black hole solutions for this action elsewhere \cite{us}. 

As a preview, we consider a perturbative treatment in the dimensionless parameter $\tilde{\alpha}\equiv \alpha/l^2$, where $l$ is the characteristic length of the system in question, e.g., the radius of the black hole horizon. Assuming $\tilde{\alpha}\ll1$ (which is a reasonable assumption unless one is considering microscopic black holes) one could look for solutions that are perturbatively close to the Schwarzschild solution. At zeroth order the scalar would then be constant. This implies that the $\phi {\cal G}$ term will only start contributing to the field equations of the metric at order ${\cal O}(\tilde{\alpha}^2)$.  Hence, to ${\cal O}(\tilde{\alpha})$ the metric will be Schwarzschild. For the scalar, instead, one can solve eq.~(\ref{eqphi2}) to ${\cal O}(\tilde{\alpha})$ and obtain
\be
\phi'=\alpha{\frac {16{m}^{2}-C {r}^{3}}{{r}^{4} \left( r-2m \right)}}    
\ee
where $m=l/2$ is the mass of a black hole and $C$ is an integration constant. For $\phi$ to be regular on the black hole horizon one must impose $C=2/m$. This yields
\bea
\phi'&=&-\frac{2 \alpha}{m}{\frac {(r^2+2m r +4m^2)}{{r}^{4} }}  \nn\\&=& -\frac{8 \tilde{\alpha} m}{r^4}(r^2+2m r +4m^2)
\eea
Two remarkable features of the solution are already present at ${\cal O}(\tilde{\alpha})$: i) even though $\phi$ has a non-trivial profile it does not lead to an independent charge because of the regularity condition on the horizon, so the solutions will have hair of the ``second kind''; ii) for fixed $\alpha$ the solution diverges as $m\to 0$. The expansion parameter is in fact $\tilde{\alpha}\propto \alpha/m^2$ and, hence, nonperturbative effects will be important in this regime. A more detailed analysis of these features and the full perturbative and non-perturbative solutions will be presented in Ref.~\cite{us}.

 The fact that the scalar field is obliged to have a non-trivial configuration in black hole spacetimes constitutes a counter-example to the statement that the most general shift-symmetric scalar-tensor theory that leads to second order field equations cannot have hairy solutions.
Indeed, the theory (\ref{gbaction}) fits comfortably in the initial Lagrangian given in Eq.~(\ref{lagr}). One simply has to choose  $K=M_p^2 X/2$, $G_3=0$, $G_4=M_p^2/2$, and $G_5=-2M_p^2\alpha \ln|X|$ \cite{Kobayashi:2011nu}. It is straightforward to check that, for these choices, the $G_5$-related terms in $J^r$ in eq.~(\ref{sphJ}) become $\phi$-independent, without the current (or any other equation of the theory) becoming divergent as $\phi\to\;$constant. It is crucial to point out that one does not need to restrict oneself to that choice in order to have hairy black holes. In fact,  for any choice of $K$ and $G_i$, one could write 
\be
G_5=-2M_p^2 \alpha \ln|X| +\tilde{G}_5(X) \,.
\ee
Additionally, the coupling between $\phi$ and ${\cal G}$ cannot be done away with by going to another conformal frame, as is the case for a coupling of the type $\phi R$. Only when $\alpha$ is tuned to zero would  $\phi=\;$constant solutions be admissible. In other words, one could add to the action (\ref{gbaction}) virtually any other term that is shift symmetric and leads to a second order contribution to the field equations and the resulting theory would evade the no-hair theorem of Ref.~\cite{Hui:2012qt}.

From a classical perspective one can always choose to set $\alpha=0$. But if one is thinking of these theories as effective field theories, then one would need a symmetry that would protect $\alpha$ from receiving radiative corrections. For a real scalar, there are not many choices of internal symmetries. Given that the corresponding term is odd in copies of $\phi$, one could invoke symmetry under $\phi\to-\phi$. This would, however, reduce the Lagrangian of Eq.~(\ref{lagr}), and thus the applicable theory space of the no-hair theorem of Ref.~\cite{Hui:2012qt}, significantly:
\begin{align}
\label{lagrsym}
{\cal L}&=K(X) + G_{4}(X)R
\nonumber\\
&~~~~~~
+G_{4X}\left[
\left(\Box\phi\right)^2-\left(\nabla_\mu\nabla_\nu\phi\right)^2\right]\,.
\end{align}

 One the other hand, it is straightforward to extend the no-hair argument of Ref.~\cite{Hui:2012qt} to slowly rotating solutions, when it is valid in spherical symmetry. The most general stationary, axially symmetric, slowly rotating solution can take the form \cite{hartle67}
\begin{eqnarray}
\label{metric}
ds^2&=&-R(r)dt^2+S(r)d\rho^2+r^2(d\theta^2+\sin^2\theta d\varphi^2)\nonumber\\
&&+\epsilon 
r^2 \sin^2\theta \,\Omega(r,\theta) dtd\varphi+{\cal O}(\epsilon^2)\,,
\end{eqnarray}
where $R(r)$ and $S(r)$ correspond to the spherically symmetric solution, $\Omega(r,\theta)$ is a function to be determined, and $\epsilon$ is the bookkeeping parameter for the slow rotation. The key argument for arriving at this metric is  that the system should be invariant under reversal of the direction of rotation together with either $t\to -t$ or $\varphi \to -\varphi$. 

Let us now apply the same requirement to the scalar field $\phi$. Assuming it respects the symmetries of the metric in our coordinate system,  the scalar field will not depend on $t$ or $\varphi$  to all orders, {\em i.e.}~$\phi=\phi(r,\theta)$. But then the scalar cannot receive a correction which is linear in the rotation,  as the linear correction would not be invariant under the combined operation mentioned above. Given that $\phi=$constant  in the spherical case, we will then have $\phi=\;$constant$+{\cal O}(\epsilon^2)$ in the slowly rotating case. The metric will then satisfy Einstein's equation to the same order. Therefore, slowly rotating black holes cannot have scalar hair.
This extension to the proof of Ref.~\cite{Hui:2012qt} is valid when a perturbative treatment in the rotation is applicable. It is a stronger result, in the sense that it demonstrates that moderate rotation cannot endow the black hole with scalar hair.  Additionally, this simple argument applies to virtually any gravity theory with scalar fields, as long as the spherically symmetric solutions have constant profiles for the scalars.

In summary, we have shown that  in generalised scalar-tensor theories that are shift-symmetric and lead to second order equations the scalar field will have a nontrivial configuration in any spacetime other than flat, unless the linear coupling between the scalar and the Gauss-Bonnet invariant is suppressed. In the absence of a symmetry justifying such suppression, black holes will be endowed with scalar hair. On the other hand, we have also argued that, when it is valid to assume that static, spherically symmetric black holes will have no hair, their slowly rotating counterparts will not have hair either.

Some comments are in order before closing. Firstly, we use the term ``hair'' loosely, to mean that the scalar has a nontrivial configuration in the black hole spacetime. This does not necessarily imply that the black hole has to carry some independent scalar charge (indeed it will not in the case of the action (\ref{gbaction}) \cite{us}). It is, however, enough to argue that the black hole will be different than its general relativity counterpart. Secondly, our attention has been focussed on shift-symmetric theories because in specific examples of scalar-tensor theories where the scalar does not exhibit shift symmetry black holes with hair are already known, see for example Ref.~\cite{Kanti:1995vq}. 

It is also worth mentioning that one could contest two more of the assumptions of any no-hair theorem for scalar fields. The first one is that the scalar has to respect the symmetries of the metric. This might be particularly relevant in the context of the shift-symmetric theories considered here, because the scalar field appears in the field equations only through its derivatives. Hence,  if one is interested in a spacetime where ${\cal L}_\xi g_{\mu\nu}=0$, where ${\cal L}_\xi$ is the Lie derivative along the generator of the symmetry $\xi$, 
it suffices to impose ${\cal L}_\xi \nabla_\mu \phi=0$. This is a weaker condition than imposing that ${\cal L}_\xi \phi=0$, as is usually done.  It is not clear, however, if such solutions will be physically relevant.

It is also important to consider whether hair can be induced by the presence of matter in the vicinity of the black hole or by embedding the black hole in a cosmological background. In standard scalar-tensor theories, both cases lead to generation of scalar hair \cite{Jacobson:1999vr,Horbatsch:2011ye,Cardoso:2013fwa,Cardoso:2013opa,Berti:2013gfa,Rinaldi:2012vy}.

{\bf\em Acknowledgements:} The authors would like to thank Enrico Barausse, Paolo Pani and Ian Vega for a critical reading of this manuscript.  The research leading to these results has received funding from the European Research Council
under the European Union's Seventh Framework Programme (FP7/2007-2013) / ERC
Grant Agreement n. 306425 ``Challenging General Relativity".



\end{document}